\documentclass[letterpaper]{article} 
\usepackage{aaai2026}  
\usepackage{times}  
\usepackage{helvet}  
\usepackage{courier}  
\usepackage[hyphens]{url}  
\usepackage{graphicx} 
\usepackage{natbib}  
\usepackage{caption} 
\usepackage{algorithm}
\usepackage{algorithmic}

\usepackage{subcaption}
\usepackage{amssymb} 
\usepackage[table]{xcolor} 
\usepackage{colortbl}      
\usepackage{amsmath}
\usepackage{float} 
\usepackage{multirow}
\usepackage{array} 
\usepackage{booktabs}
\usepackage{amsmath}
\usepackage{amssymb}
\usepackage{enumitem}
\usepackage{physics}
\usepackage{bm} 

\newcommand{\vect}[1]{\boldsymbol{#1}}  
\usepackage[table]{xcolor} 

\usepackage{newfloat}
\usepackage{listings}
\DeclareCaptionStyle{ruled}{labelfont=normalfont,labelsep=colon,strut=off} 
\floatstyle{ruled}
\newfloat{listing}{tb}{lst}{}
\floatname{listing}{Listing}
\title{DynaQuant: Dynamic Mixed-Precision Quantization \\ for Learned Image Compression}
\author {
    Youneng Bao\textsuperscript{\rm 1},
    Yulong Cheng\textsuperscript{\rm 2}, 
    Yiping Liu\textsuperscript{\rm 2}, 
    Yichen Yang\textsuperscript{\rm 3}, 
    Peng Qin\textsuperscript{\rm 4}, \\
    Mu Li\textsuperscript{\rm 2,*}, 
    Yongsheng Liang\textsuperscript{\rm 1,2,}\thanks{Corresponding author}
}
\affiliations {
    \textsuperscript{\rm 1}College of Electronics and Information Engineering, Shenzhen University\\
    \textsuperscript{\rm 2}Harbin Institute of Technology, Shenzhen\\
    \textsuperscript{\rm 3}Marine Design and Research Institute of China\\
    \textsuperscript{\rm 4}China Telecom Group Qinhuangdao Branch\\
    baoyouneng@163.com, \{23s151073, yipingliu\}@stu.hit.edu.cn, 
    yicen1994@163.com, \\
    2244626474@qq.com,
    \{limu2022, liangys\}@hit.edu.cn
}

\usepackage{bibentry}

\begin{document}

\maketitle

\begin{abstract}
Prevailing quantization techniques in Learned Image Compression (LIC) typically employ a static, uniform bit-width across all layers, failing to adapt to the highly diverse data distributions and sensitivity characteristics inherent in LIC models. This leads to a suboptimal trade-off between performance and efficiency. In this paper, we introduce DynaQuant, a novel framework for dynamic mixed-precision quantization that operates on two complementary levels. First, we propose content-aware quantization, where learnable scaling and offset parameters dynamically adapt to the statistical variations of latent features. This fine-grained adaptation is trained end-to-end using a novel Distance-aware Gradient Modulator (DGM), which provides a more informative learning signal than the standard Straight-Through Estimator. Second, we introduce a data-driven, dynamic bit-width selector that learns to assign an optimal bit precision to each layer, dynamically reconfiguring the network's precision profile based on the input data. Our fully dynamic approach offers substantial flexibility in balancing rate-distortion (R-D) performance and computational cost. Experiments demonstrate that DynaQuant achieves rd performance comparable to full-precision models while significantly reducing computational and storage requirements, thereby enabling the practical deployment of advanced LIC on diverse hardware platforms.

\end{abstract}

\begin{links}
    \link{Code}{https://github.com/baoyu2020/DynaQuant}
\end{links}

\section{Introduction}
\label{sec:introduction}
Learned Image Compression (LIC) has emerged as a new paradigm in image coding, achieving remarkable success in recent years. By replacing traditional hand-crafted modules—such as transforms, quantization, and entropy coding—with end-to-end optimized deep neural networks, LIC models~\cite{ballé2018variational,he2022elic, BaoMLMTL23,10.1609/aaai.v37i1.25184,zeng2025mambaic} have consistently surpassed the R-D performance of state-of-the-art conventional codecs like BPG~\cite{sullivan2012overview} and VVC~\cite{bross2021overview}. This data-driven methodology offers unprecedented flexibility and performance, marking a significant milestone in image coding.

Despite their impressive R-D performance, the practical deployment of these advanced LIC models is severely hampered by their substantial computational complexity and large memory footprint. The deep, complex architectures, while powerful, are computationally intensive, rendering them ill-suited for real-time applications on resource-constrained hardware such as smartphones, drones, and other edge devices. 

\begin{figure}[t]
    \centering    \includegraphics[width=1\linewidth]{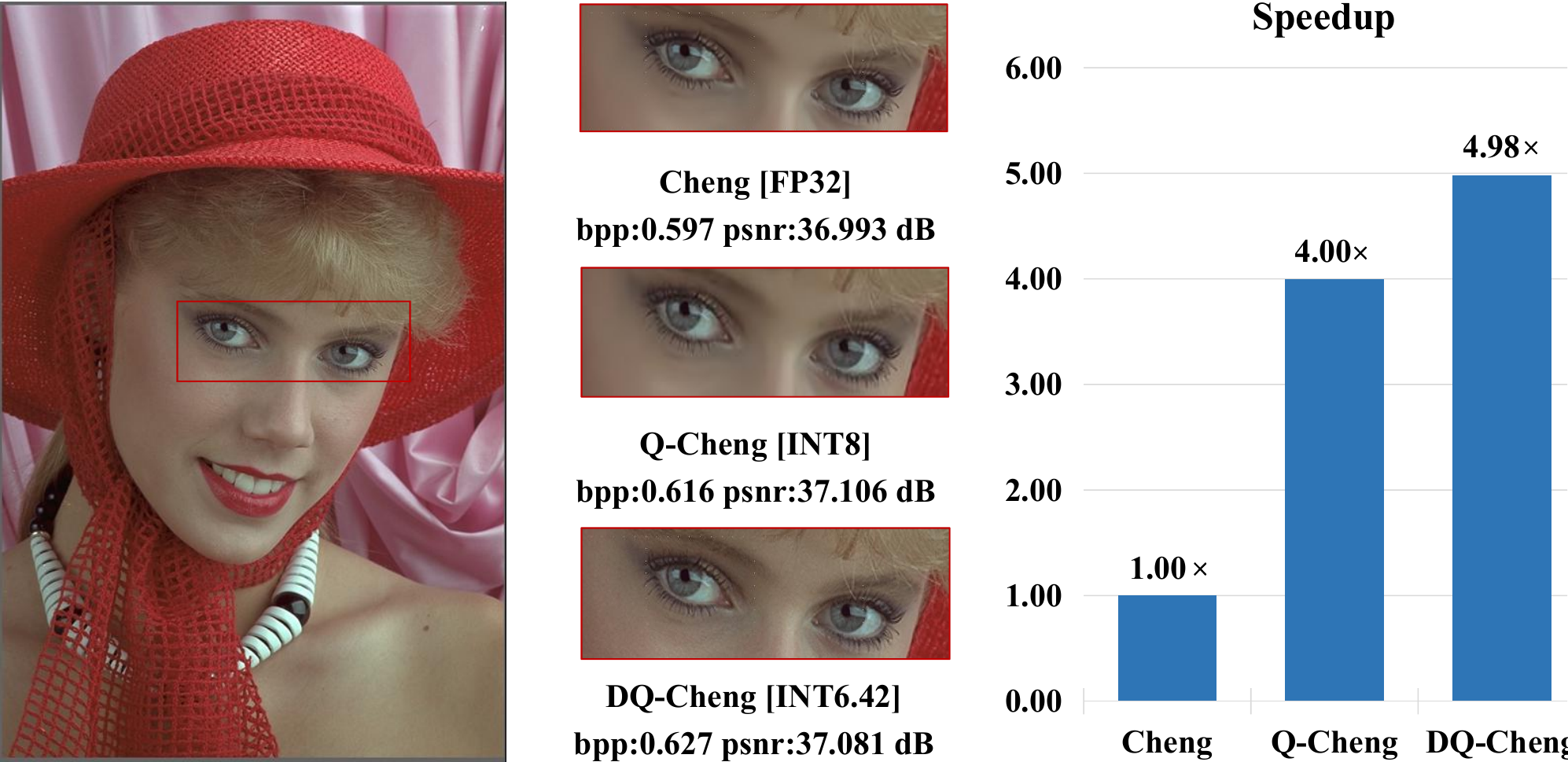}
    \caption{Visual and quantitative comparison of the DynaQuant method on the \textit{kodim04}, using {Cheng2020}~\cite{cheng2020learned} as the baseline. Our proposed two quantization strategies—fixed bit-width quantization (Q-Cheng) and dynamic bit-width quantization (DQ-Cheng)—achieve comparable performance to the full-precision method while delivering approximately {5 $\times$ speedup}.}
    \label{fig:vis_kodim04}
\end{figure}

To bridge this gap, model quantization has become a standard technique for compressing and accelerating neural networks. However, most existing quantization methods, often developed for robust high-level vision tasks like image classification, are suboptimal for the high-fidelity demands of image compression. These methods typically apply a static, uniform bit-width across all network layers. This ``one-size-fits-all" approach fundamentally ignores two critical properties of LIC models: (1) Content-dependent dynamics: The statistical distributions of latent representations in LIC models are highly non-stationary and vary significantly with the content of each input image. A fixed set of quantization parameters cannot adapt to these dynamic distributions, leading to significant information loss. (2) Layer-wise sensitivity: Different layers within an LIC model exhibit vastly different sensitivities to quantization noise. Forcing a uniform precision profile is inefficient, as it may under-quantize robust layers while over-quantizing sensitive ones, resulting in a poor trade-off between accuracy and complexity.

To address these limitations, we propose DynaQuant, a framework for dynamic mixed-precision quantization tailored specifically for LIC. Our key insight is that an optimal quantization strategy must be dynamic at two complementary levels: the parameter level and the architectural level. First, to handle content-dependent data distributions, inspired by the soft to hard method~\cite{DBLP:conf/nips/AgustssonMTCTBG17},  we introduce content-aware quantization, where the quantization parameters are learned and dynamically adjusted for each input. This process is guided by a novel distance-aware gradient modulator, which provides a more meaningful learning signal for the non-differentiable rounding operation than the conventional STE. Second, to address varying layer sensitivities, we design a lightweight, data-driven dynamic bit-width selector. This module learns to assign an optimal bit-width to each layer on-the-fly, creating an input-specific precision profile for the network architecture. This dual-level dynamic adaptation allows for an unprecedentedly fine-grained balance between rate-distortion performance and computational efficiency.
Our main contributions can be summarized as follows:
\begin{itemize}
    \item  A novel dynamic mixed-precision quantization framework, DynaQuant, for LIC that adapts both quantization parameters and layer-wise bit-widths in a content-aware manner.
    \item A differentiable dynamic bit-width selector that learns to allocate optimal precision to each network layer by jointly optimizing the R-D loss of LIC with differentiable mixed-precision allocation.
    \item  An extensive experiment demonstrating that DynaQuant achieves R-D performance on par with full-precision models while significantly reducing computational complexity and model size, thereby enabling practical deployment of LIC on diverse hardware platforms.
\end{itemize}

\section{Related Work}
\label{sec:related}

\subsection{Learned Image Compression}

LIC has surpassed traditional codecs like JPEG~\cite{wallace1991jpeg},  JPEG2000~\cite{skodras2002jpeg} and VVC~\cite{bross2021overview} in rate-distortion performance. This was pioneered by end-to-end trained autoencoder architectures~\cite{ballé2018variational}. Subsequent works have improved these architectural designs by introducing advanced network structures and attention mechanisms to enhance representation learning and model long-range dependencies~\cite{TanMLBL25,bai2021endtoendimagecompressionanalysis,Zheng_Gao_2024}. Recent methods leverage multi-modality networks with semantic priors~\cite{10.1609/aaai.v37i1.25184} or replace CNNs with Transformer~\cite{lu2021transformer,liu2023learned,li2023frequency} and Mamba~\cite{qin2024mambavc,wu2025cmamba} for improved global modeling while preserving local detail~\cite{liu2023learned,zeng2025mambaic}. 
In parallel, advances in entropy and context modeling have improved LIC performance, such as multi-reference entropy models for more accurate probability estimation~\cite{jiang2023mlic, BaoTJLLT25}. However, these collective advances in network architecture and entropy modeling substantially increase computational complexity, limiting practical deployment on resource-constrained devices.

To reduce the high computational cost of LIC, various model compression techniques have been explored. Pruning methods reduce model parameters via regularization-based sparsification~\cite{chen2023efficient} or structured removal of network components~\cite{kim2020efficient,luo2022memory}. An alternative paradigm fits lightweight models to individual images, thereby simplifying the entropy model and feature extractors~\cite{kim2024c3,balle2025good,zhang2025fitted}. 

Parameter quantization is also widely used, employing strategies like layer-wise or low-bit quantization to reduce model size and accelerate inference~\cite{hong2020efficient,jeon2023integer}. Among these techniques, quantization is particularly suitable for practical deployment due to its implementation simplicity and hardware support. 

However, existing quantization methods in LIC are typically static. They often apply uniform bit-widths and treat model components in isolation, an approach that lacks joint dynamic optimization. Consequently, these methods yield suboptimal R-D trade-offs and fail to fully exploit the potential of quantization.

\subsection{Model Quantization}
Model quantization is typically divided into two categories: Post-Training Quantization (PTQ) and Quantization-Aware Training (QAT).

\textbf{PTQ} quantizes models post-training without retraining but often suffers severe performance degradation at ultra-low bit-widths. Various methods have been proposed to mitigate this issue: data-adaptive rounding via quadratic optimization and continuous relaxation~\cite{nagel2020up}, random dropping of quantization during inference to maintain accuracy~\cite{wei2022qdrop}, weight quantization guided by second-order information~\cite{frantar2023optimalbraincompressionframework}, and direct optimization of the rate-distortion loss for task-oriented image compression~\cite{shi2023rate}.

\textbf{QAT} integrates quantization into the training loop, learning parameters via backpropagation to mitigate accuracy loss. Its development centers on two core challenges: devising adaptive quantization strategies and ensuring training stability. Static parameters struggle with data and layer diversity, while quantization's discrete nature destabilizes gradients. Solutions include learnable step sizes for data adaptation~\cite{esser2019learned}, quantization interval learning for stability~\cite{DBLP:journals/corr/abs-1808-05779}, and content-adaptive methods that dynamically adjust parameters to each input~\cite{liu2022instance}.

\subsection{Mixed Precision Quantization} 

Mixed-precision quantization assigns different bit-widths to different layers, mitigating the performance loss of uniform-precision methods. This allocation can be modeled as a search problem, solved using reinforcement learning~\cite{wang2019haq} or search-based methods at kernel-level granularity~\cite{DBLP:journals/corr/abs-1902-05690}. Alternatively, bit-widths can be determined based on layer sensitivity, with some methods estimating this sensitivity through second-order Hessian information~\cite{DBLP:journals/corr/abs-1911-03852,dong2019hawqhessianawarequantization}. Other works~\cite{liu2019dartsdifferentiablearchitecturesearch} formulate bit-width allocation as a differentiable optimization problem, enabling layer precisions to be learned directly via gradient descent. In LIC, some approaches guide bit-width allocation by analyzing feature entropy~\cite{sun2022entropy}. Others~\cite{hossain2024flexible} have proposed flexible mixed-precision frameworks that provide customizable solutions for LIC models.

To address the limitation of employing a static, uniform bit-width across all layers, we propose a bit-selector with key advantages: (1) minimal computational overhead; (2) direct rate-distortion optimization. To our knowledge, this is the first end-to-end image compression method that combines trainable quantization parameters with differentiable mixed-precision allocation.

\section{Proposed Method}
\label{sec:method}

\subsection{Preliminaries: Quantization-Aware Training}
QAT simulates the effects of quantization during training, allowing the model to adapt to precision loss. For a given full-precision input $x$, a $b$-bit asymmetric quantizer first maps it to an integer $x_q$ using a scale factor $s$ and a zero-point $z$:
\begin{equation}
x_q = \text{round}\left(\text{clip}\left(\frac{x}{s} + z, n_{\min}, n_{\max}\right)\right),
\label{eq:quant}
\end{equation}
where $[n_{\min}, n_{\max}]$ defines the quantization range (e.g., $[0, 2^b-1]$ for unsigned $b$-bit integers). The scale $s$ and zero-point $z$ are typically pre-computed from the activation statistics. The de-quantized value $\tilde{x}$, which approximates the original input $x$, is recovered as:
\begin{equation}
\tilde{x} = s \cdot (x_q - z).
\label{eq:dequant}
\end{equation}

\noindent In conventional QAT, the scale $s$ and zero-point $z$ are statically computed from aggregated activation statistics and remain fixed during the inference stage.

To enable gradient-based optimization, the non-differentiable $\textit{round}$ function  is commonly handled by the STE, which approximates its gradient as an identity function:
\begin{equation}
\frac{\partial \mathcal{L}}{\partial x} \approx \frac{\partial \mathcal{L}}{\partial \tilde{x}} \quad \text{if } x \in [s(n_{\min}-z), s(n_{\max}-z)].
\label{eq:ste}
\end{equation}

\begin{figure}[ht!]
    \centering
    \includegraphics[width=1\linewidth]{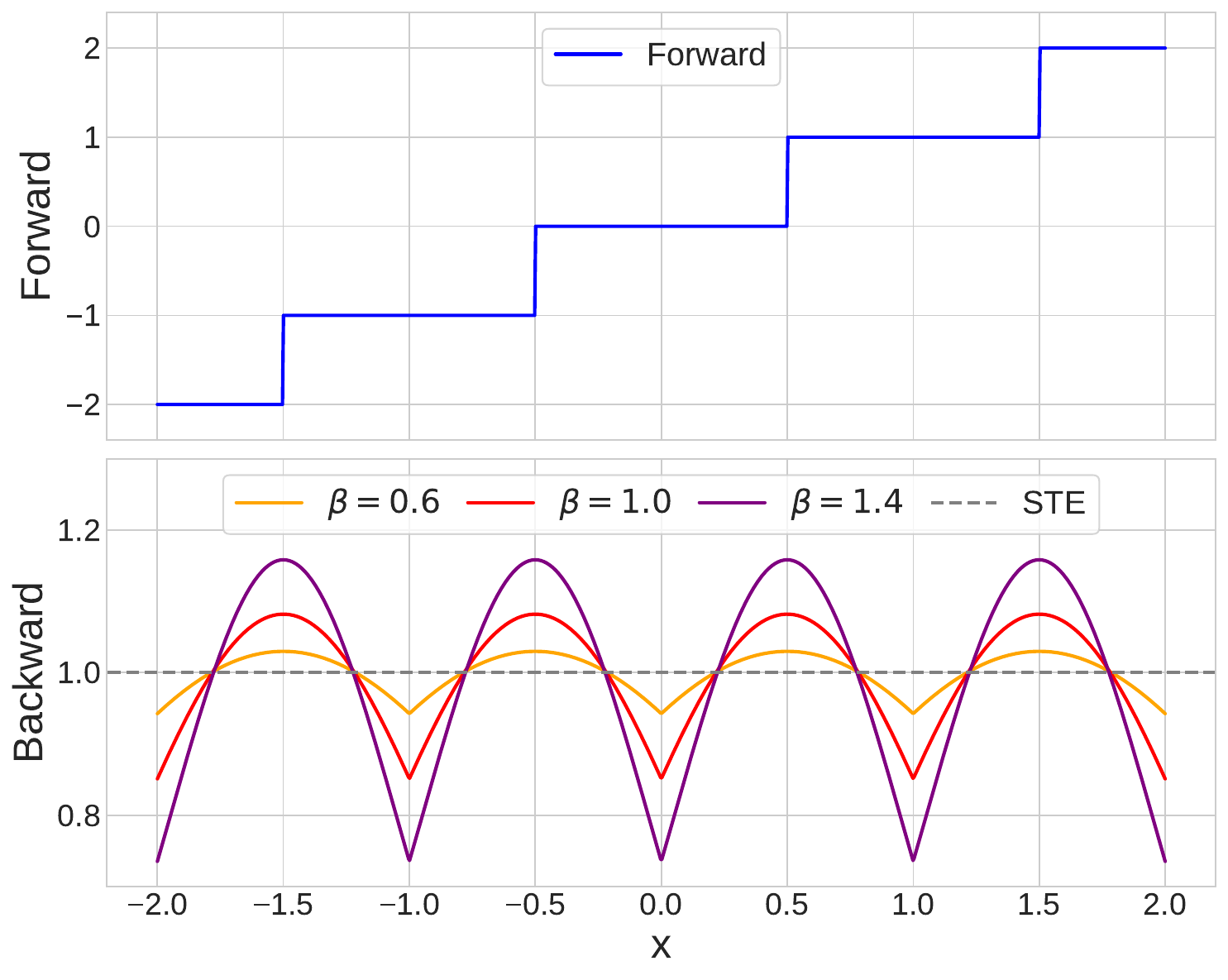}
    \caption{Gradient proxy function (top) and its derivative (bottom). The derivative exhibits periodic oscillations, reaching minima at $x$=0 and $x$=1 with peaks at $x$=0.5. All values remain strictly positive, and the amplitude is modulated by $\beta$, ensuring adaptive gradient scaling.}
    \label{fig:proxy}
\end{figure}

However, these two cornerstones of standard QAT—static parameters and a coarse, uniform gradient—are fundamentally ill-suited for the high-fidelity demands of LIC. Static parameters can not adapt to content-dependent variations in latent distributions, while the STE's uniform gradient signal fails to provide meaningful feedback for optimizing the quantization process itself. 

\subsection{Dynamic Parameter Adaptation (DPA)}

To overcome the rigidity of static quantization, we introduce a mechanism that dynamically adapts quantization parameters to the specific characteristics of each input, combining content-aware parameter learning with a more principled gradient approximation.

\subsubsection{Content-Aware Quantization Mapping}

The latent representations in LIC models exhibit highly dynamic and non-uniform distributions that depend on image content. To adapt to these distributions, we eschew fixed quantization parameters. Instead, we define the scale factor $s$ and zero-point $z$ as learnable, per-channel parameters. This allows the network to learn an optimal quantization mapping for each feature map, minimizing rate-distortion loss. The forward pass follows the standard asymmetric quantization-dequantization process:
\begin{equation}
\tilde{x} = s \cdot \left(\text{round}\left(\text{clip}\left(\frac{x}{s} + z, n_{\min}, n_{\max}\right)\right) - z\right),
\label{eq:forward_pass}
\end{equation}
where the clipping range $[n_{\min}, n_{\max}]$ is set to $[0, 2^b-1]$ for a $b$-bit quantizer. The main challenge, however, lies in optimizing $s$ and $z$ effectively, which requires a valid gradient through the non-differentiable rounding function.

\begin{figure*}[t]
    \centering    \includegraphics[width=0.8\linewidth]{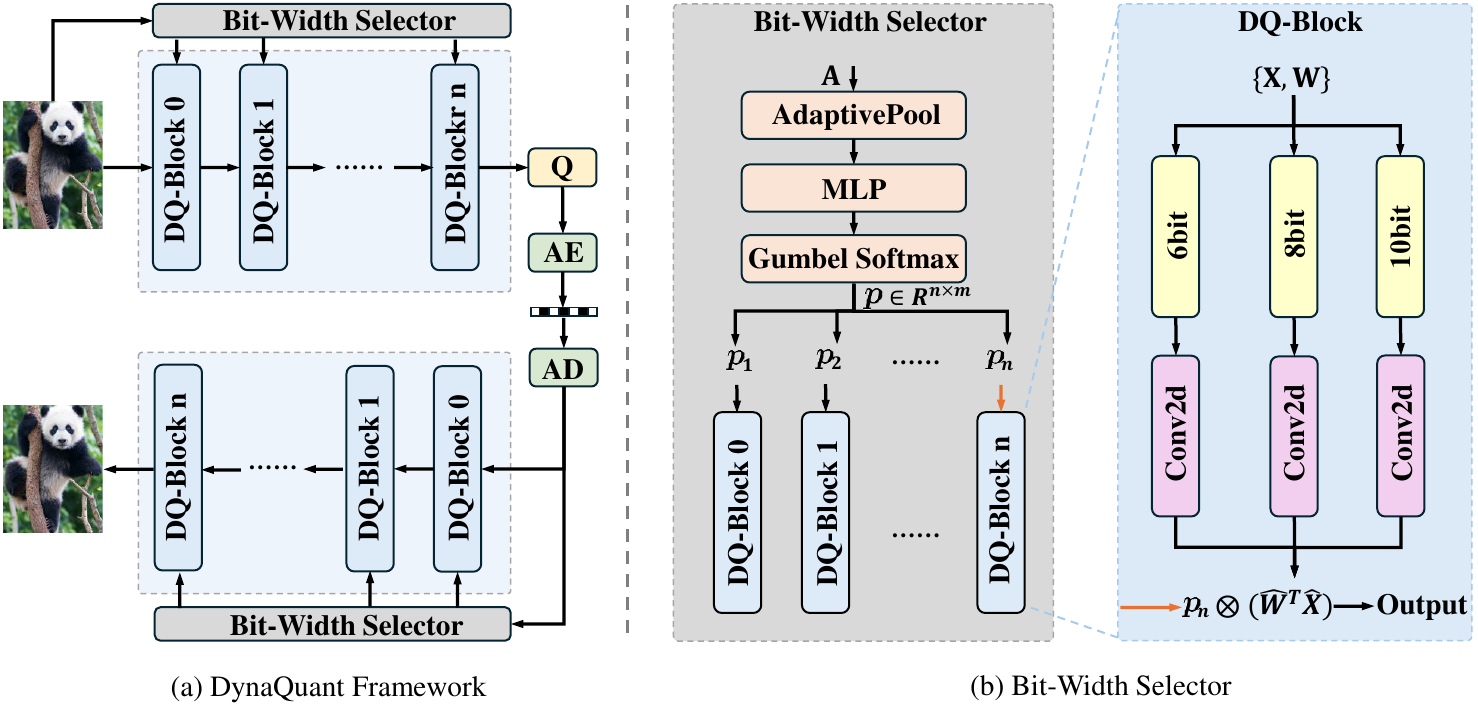}    
    \caption{(a) DynaQuant Framework Overview. DQ-Block is the dynamic quantization block, and Bit-Width Selector is the bit-width selector that dynamically allocates quantization precision for each layer. (b) Bit-Width Selector and DQ-Block Structure. The bit-width selector processes input activation $A$ through adaptive pooling, MLP, and Gumbel Softmax to output bit-width selection probability distribution $p_1, p_2, \ldots, p_n$. DQ-Block quantizes the input $\{X\}$ and learnable parameters $\{W\}$ within the module according to the corresponding bit-widths based on the probability distribution, and finally generates the output through probability-weighted fusion.}
    \label{fig:framework}
\end{figure*}

\subsubsection{Distance-Aware Gradient Modulation (DGM)}
The Straight-Through Estimator, which approximates the rounding gradient as a constant $1$, is the de facto standard. However, it implicitly assumes that quantization error is uniformly sensitive to changes in the input, an assumption that does not hold. The error is most sensitive near the quantization decision boundaries (i.e., half-integer values like $0.5$).

To provide a more informative learning signal, we introduce a Distance-Aware Gradient Modulation (DGM) mechanism. The core principle of DGM is to make the magnitude of the backpropagated gradient a function of the input's proximity to the nearest decision boundary. Inspired by the work~\cite{qin2023quantsr} on model parameter quantization, we define a new gradient proxy $g'(x)$ to implement DGM:
\begin{equation}
\frac{\partial \text{round}(x)}{\partial x} \triangleq g'(x) = f(\text{dist}(x, \text{boundary})).
\end{equation}
Specifically, we formulate $g(x)$ as:
\begin{equation}
g(x) = \frac{1}{2} \cdot \frac{\tanh(\beta(x - \lfloor x \rfloor) - 0.5)}{\tanh(0.5)} + 0.5 \label{eq:dgm_complete}
\end{equation}
where $x$ is the value before rounding and $\beta$ is a shape factor. As illustrated in~Fig.~\ref{fig:proxy}, this function yields a gradient that gradually decays as the input moves from decision boundaries toward quantization centers. Consequently, parameters associated with uncertain, boundary-adjacent values receive larger updates, while those associated with stable values receive smaller updates. This targeted gradient modulation encourages the model to learn better quantization parameters, enabling content-adaptive quantization and improving overall quantization performance.

\begin{figure*}    
  \centering
  \begin{subfigure}[b]{0.32\textwidth}
    \centering
    \includegraphics[width=\linewidth]{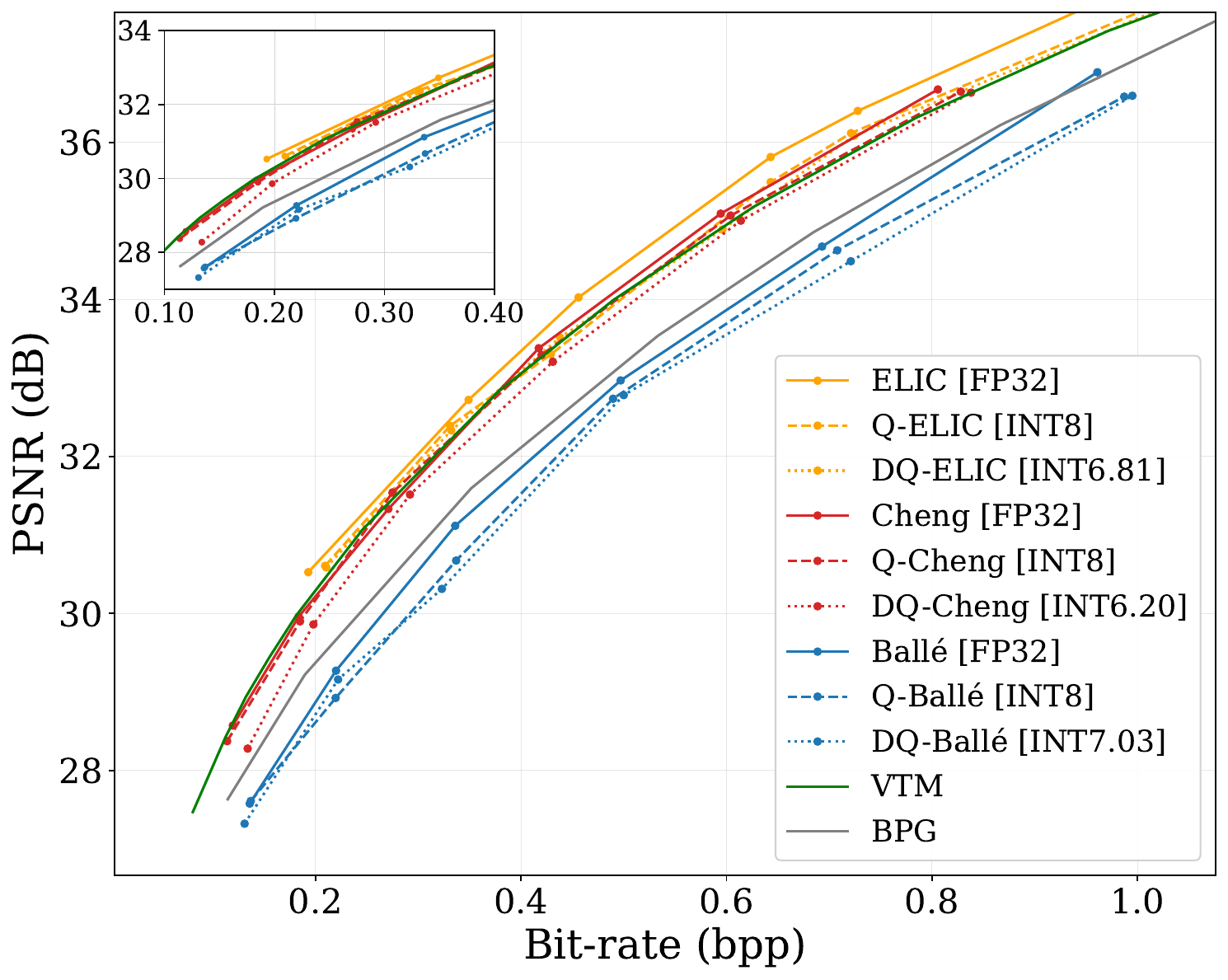}
    \caption{}
    \label{subfig:a}
  \end{subfigure}
  \hfill
  \begin{subfigure}[b]{0.32\textwidth}
    \centering
    \includegraphics[width=\linewidth]{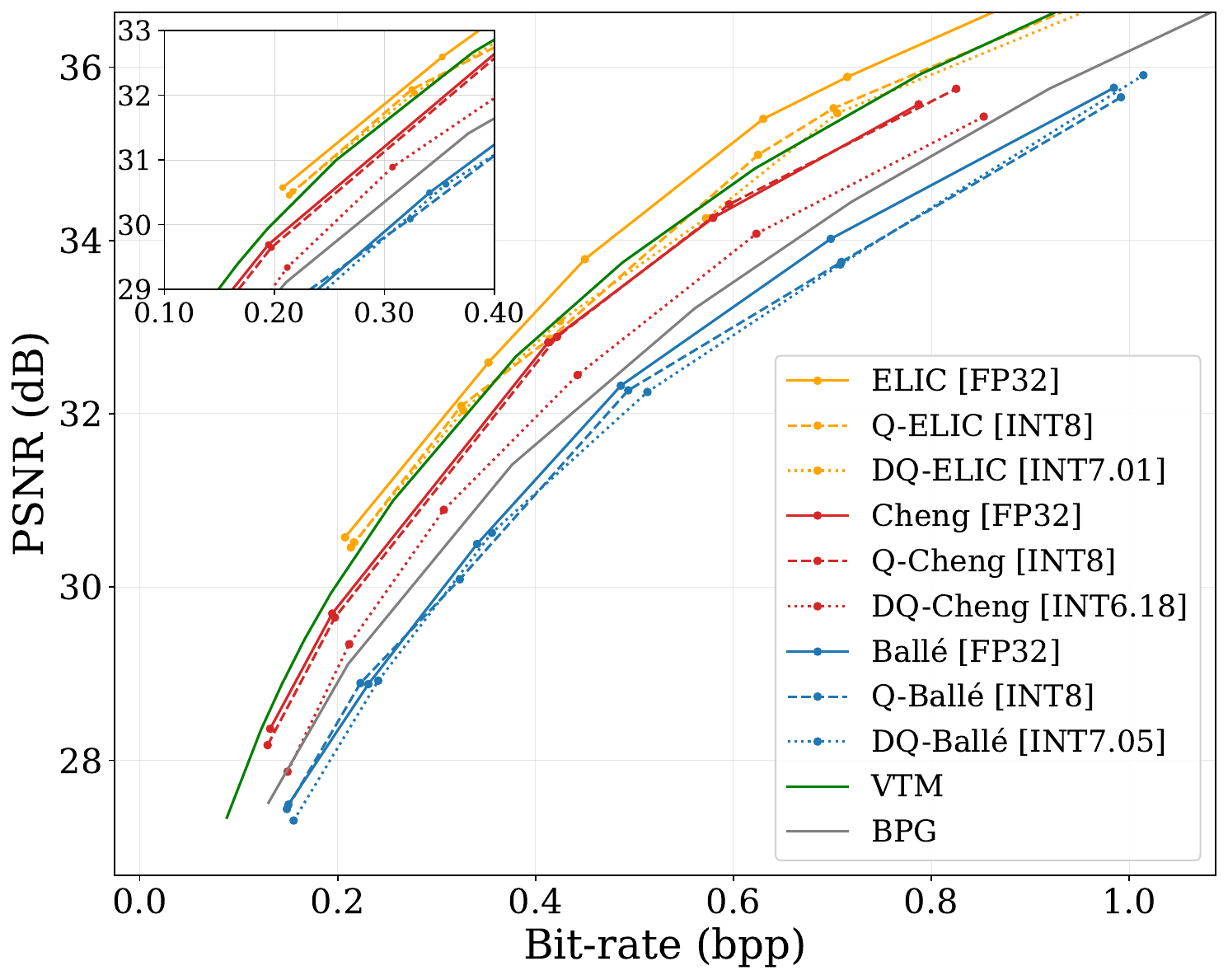}
    \caption{}
    \label{subfig:b}
  \end{subfigure}
  \hfill
  \begin{subfigure}[b]{0.32\textwidth}
    \centering
    \includegraphics[width=\linewidth]{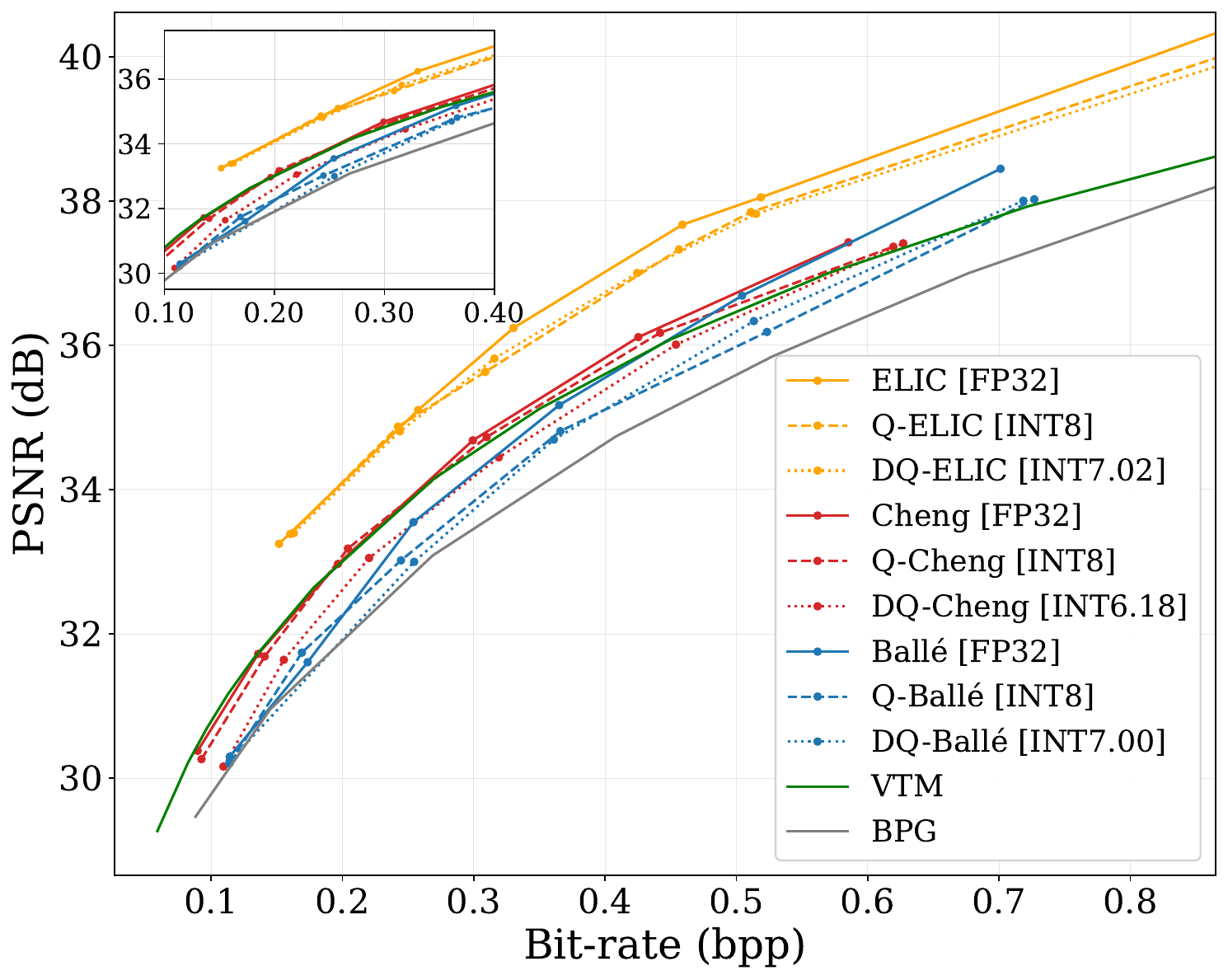}
    \caption{}
    \label{subfig:c}
  \end{subfigure}  
  \caption{{\bf R-D Performance.}(a) Kodak; (b) JPEG-AI; (c) CLIC. Quantization schemes: [FP32] for original 32-bit float, [INT8] for uniform 8-bit quantization, and [INTX.YY] for our mixed-precision quantization where X.YY indicates the average bit-width (e.g., 6.81-bit for DQ-ELIC, 6.20-bit for DQ-Cheng). ``Q-ELIC" refers to applying DPA to each layer of ELIC, while ``DQ-ELIC" denotes the application of a DBWS to each layer of ELIC. Best viewed in color.}
  \label{fig:sota_rd_curves}
\end{figure*}

\subsection{Dynamic Bit-Width Selector (DBWS)}
Beyond adapting parameters, we introduce a method for layer-wise adaptive bit allocation. Instead of relying on costly search algorithms, we design a lightweight, data-driven module that learns to assign an optimal bit-width to each layer in an end-to-end fashion.

\subsubsection{Content-Aware Bit-Width Selector }

We propose a differentiable bit-width selector, a small network module that predicts the optimal bit-width for a given layer based on its activation statistics. The selector takes the input activation tensor $A \in \mathbb{R}^{C \times H \times W}$ and performs the following three steps:

\begin{enumerate}
    \item \textbf{Global Feature Extraction:} An adaptive pooling layer (e.g., {AdaptivePool}) first condenses the spatial dimensions of the activation tensor into a fixed-size feature vector $\vect{h}_{\text{pool}} = \text{AdaptivePool}(A)$. This vector captures the global statistical properties of the activations while remaining agnostic to the input tensor's dimensions.

    \item \textbf{Bit-Width Prediction:} The feature vector is then processed by a two-layer MLP to predict a probability distribution over a set of candidate bit-widths $\mathcal{B} = \{b_1, b_2, \ldots, b_M\}$.
    \begin{equation}
        \bm{p} = \text{Softmax}(\text{MLP}(\vect{h}_{\text{pool}}))
    \end{equation}
    where $\bm{p} = \{p_1, p_2, \ldots, p_N\}$. $N$ is the total number of quantizable blocks, for each quantizable block $l$. $\bm{p}_l \in \mathbb{R}^M$ is the probability vector, with $(\bm{p}_l)_k$ representing the probability of selecting bit-width $b_k$.

    \item \textbf{Stochastic Selection for Training:} To make the discrete selection process differentiable, we employ the Gumbel-Softmax reparameterization trick during training. This allows us to sample a one-hot vector $\vect{p}_l$ from the categorical distribution defined by $\bm{p}$ while allowing gradients to flow back to the selector's parameters. The effective bit-width for the layer is then calculated as $b_l = \sum_{k=1}^{M} (\vect{p}_l)_k \cdot b_k$.
\end{enumerate}

During inference, we deterministically select the bit-width with the highest probability to ensure stable and efficient execution: $b_l^* = b_k$ where $k = \text{argmax}(\bm{p}_l)$. This mechanism allows each layer to dynamically choose its precision.

\subsection{Joint Optimization Framework}

Our framework is trained end-to-end by minimizing a composite loss function that jointly optimizes for rate-distortion performance and model complexity. The overall learning objective $\mathcal{L}$ is defined as:
\begin{equation}
    \mathcal{L} = R + \lambda D + \gamma \mathcal{L}_{\text{bits}},
    \label{eq:total_loss}
\end{equation}
where $R$ is the estimated bitrate of the quantized latent representation, derived from the entropy model. $D$ measures the distortion between the original image and the reconstructed image (e.g., using MSE or MS-SSIM); $\mathcal{L}_{\text{bits}}$ is a complexity regularization term, and $\lambda$ is a hyperparameter that balances the two objectives. The hyperparameter $ \gamma$ is a hyperparameter that balances
the R-D performance and bit-width.

The core of our adaptive training is the complexity regularization term, $\mathcal{L}_{\text{bits}}$, which guides the bit-width selector. To encourage the network to favor lower-precision configurations, we define this term as the expected average bit-width across all $L$ dynamically quantized layers:
\begin{equation}
    \mathcal{L}_{\text{bits}} = \frac{1}{L} \sum_{l=1}^{L} \mathbb{E}_{\bm{p}_l}[b] = \frac{1}{L} \sum_{l=1}^{L} \sum_{k=1}^{M} (\bm{p}_l)_k \cdot b_k,    \label{eq:bit_loss_standard}
\end{equation}
where $(\bm{p}_l)_k$ is the probability of selecting the $k$-th candidate bit-width $b_k \in \mathcal{B}$ for layer $l$. This loss term creates a direct optimization pressure on the bit-width selection policy, rewarding it for reducing the average bit-width. By adjusting $\gamma$, we can effectively navigate the trade-off between model performance and its computational/storage footprint, producing models that are optimized for various efficiency constraints.

\section{Experiments}
\label{sec:experiments}
\begin{table*}  
  \setlength{\tabcolsep}{2.5pt} 
\centering
\caption{BD-Rate loss (\%) and computational speed up ($\times$) relative to baseline 32-bit full precision models on three standard datasets: Kodak, JPEG-AI, and CLIC. ``Q-ELIC" refers to applying Dynamic Parameter Adaptation (DPA) to each layer of ELIC, while ``DQ-ELIC" denotes the application of a Dynamic Bit-Width Selector (DBWS) to each layer of ELIC. A dash``--'' denotes unreported results; the asterisk ``*'' indicates our calculated results.
}
\label{tab:rd_performance}
\begin{tabular}{llcccccccccc}
\toprule
\multirow{2}{*}{Model} & \multirow{2}{*}{Method} & \multicolumn{2}{c}{Kodak} & \multicolumn{2}{c}{JPEG-AI} & \multicolumn{2}{c}{CLIC} & \multicolumn{4}{c}{Average} \\
\cmidrule(lr){3-4} \cmidrule(lr){5-6} \cmidrule(lr){7-8} \cmidrule{9-12}
 & & {BD-Rate} & {Speedup} & {BD-Rate} & {Speedup} & {BD-Rate} & {Speedup} &{Bitwidth} &{BD-Rate} & {Speedup} &{Model Size} \\
\midrule
\multirow{3}{*}{ELIC} & Full precision  & 0.00 & 1.00$\times$ & 0.00 & 1.00$\times$ & 0.00 & 1.00$\times$ & 32.00 & 0.00 &1.00$\times$ &137.11 MB\\
&\cellcolor{gray!30}Q-ELIC & \cellcolor{gray!30} 5.97 & \cellcolor{gray!30} 4.00$\times$ & \cellcolor{gray!30} 6.24 & \cellcolor{gray!30} 4.00$\times$ & \cellcolor{gray!30} 2.55 & \cellcolor{gray!30} 4.00$\times$ & \cellcolor{gray!30} 8.00 & \cellcolor{gray!30} 4.92 & \cellcolor{gray!30} 4.00$\times$ &\cellcolor{gray!30}34.28 MB\\
 &\cellcolor{gray!30}DQ-ELIC & \cellcolor{gray!30} 7.62 & \cellcolor{gray!30} 4.70$\times$ & \cellcolor{gray!30} 7.53 & \cellcolor{gray!30} 4.56$\times$ & \cellcolor{gray!30} 4.01 & \cellcolor{gray!30} 4.56$\times$ & \cellcolor{gray!30} 6.95 & \cellcolor{gray!30} 6.39 & \cellcolor{gray!30} 4.61$\times$ &\cellcolor{gray!30}29.78 MB\\
\midrule
\multirow{7}{*}{Cheng} & Full precision & 0.00 & 1.00$\times$ & 0.00 & 1.00$\times$ & 0.00 & 1.00$\times$ & 32.00 & 0.00 &1.00$\times$&45.08 MB\\
& 8-bit FPQ& 2.05 & *3.98$\times$ & -- & -- & 3.54 & *3.98$\times$ & 8.00& 2.80 & *3.98$\times$ &*11.31 MB\\
& FMPQ& 0.89 & 4.00$\times$ & -- & -- & 1.70 & 4.00$\times$ & --& 1.30 & 4.00$\times$ &*11.27 MB\\
& RAQ& 27.84 & -- & -- & -- & -- & -- & -- & 27.84 & --&--\\
& RDO-PTQ& 4.88 & 4.00$\times$ & -- & -- & -- & -- & 8.00& 4.88 & 4.00$\times$ & *11.27 MB\\
& \cellcolor{gray!30}Q-Cheng & \cellcolor{gray!30} 1.02 & \cellcolor{gray!30} 4.00$\times$ & \cellcolor{gray!30} 1.18 & \cellcolor{gray!30} 4.00$\times$ & \cellcolor{gray!30} 2.61 & \cellcolor{gray!30} 4.00$\times$ & \cellcolor{gray!30} 8.00 & \cellcolor{gray!30} 1.60 & \cellcolor{gray!30} 4.00$\times$ &\cellcolor{gray!30}11.27 MB\\
 & \cellcolor{gray!30}DQ-Cheng & \cellcolor{gray!30} 7.15 & \cellcolor{gray!30} 5.16$\times$ & \cellcolor{gray!30} 16.52 & \cellcolor{gray!30} 5.17$\times$ & \cellcolor{gray!30} 12.87 & \cellcolor{gray!30} 5.18$\times$ & \cellcolor{gray!30} 6.19 & \cellcolor{gray!30} 12.18 & \cellcolor{gray!30} 5.17$\times$ &\cellcolor{gray!30}8.72 MB\\
\midrule
\multirow{5}{*}{Ballé} & Full precision & 0.00 & 1.00$\times$ & 0.00 & 1.00$\times$ & 0.00 & 1.00$\times$ & 32.00 & 0.00 &1.00$\times$&19.37 MB\\
& 8-bit FPQ& 7.44 & *3.97$\times$ & -- & -- & 8.95 & *3.97$\times$ & 8.00& 8.20 & *3.97$\times$& *4.88 MB\\
& FMPQ& 6.48 & *3.98$\times$ & -- & -- & 8.95 & *3.98$\times$ & --& 7.50 & *3.98$\times$ &*4.87 MB\\
 & \cellcolor{gray!30}Q-Ballé & \cellcolor{gray!30} 5.85 & \cellcolor{gray!30} 4.00$\times$ & \cellcolor{gray!30} 2.59 & \cellcolor{gray!30} 4.00$\times$ & \cellcolor{gray!30} 6.60 & \cellcolor{gray!30} 4.00$\times$ & \cellcolor{gray!30} 8.00 & \cellcolor{gray!30} 5.01 & \cellcolor{gray!30} 4.00$\times$ &\cellcolor{gray!30}4.84 MB\\
 & \cellcolor{gray!30}DQ-Ballé & \cellcolor{gray!30} 7.63 & \cellcolor{gray!30} 4.55$\times$ & \cellcolor{gray!30} 4.84 & \cellcolor{gray!30} 4.54$\times$ & \cellcolor{gray!30} 8.06 & \cellcolor{gray!30} 4.57$\times$ & \cellcolor{gray!30} 7.03 & \cellcolor{gray!30} 6.84 & \cellcolor{gray!30} 4.55$\times$ &\cellcolor{gray!30}4.26 MB\\
\bottomrule
\end{tabular}
\end{table*}

\subsection{Experimental Setup}
\label{subsec:exp_setup}

\paragraph{Datasets and Evaluation  Metrics}
We evaluate our method on three datasets: Kodak~\cite{Kodak_2024}, JPEG-AI~\cite{jpeg_ai_dataset} and the CLIC ~\cite{CLIC2020}. Image quality is assessed using PSNR, while compression efficiency is measured in bits per pixel (bpp). BD-Rate loss is reported to quantify the average BPP difference for the same quality.

\paragraph{Baseline LIC Model and Comparison Methods}

We adopt the {Cheng2020}\cite{cheng2020learned}, {Ballé}~\cite{ballé2018variational}, and {ELIC}~\cite{he2022elic} architectures as our baseline full-precision (FP32) LIC models, recognized for their robust performance. Our proposed DynaQuant method is compared against their full-precision versions, as well as several quantization methods: 8-bit Fixed Precision Quantization (FPQ), FMPQ~\cite{10687695}, RAQ~\cite{hong2020efficient}, and RDO-PTQ~\cite{shi2023rate}. Implementation details are provided in the Appendix.

\subsection{Main Results}
\label{subsec:main_results}

\paragraph{Rate-Distortion Performance}
Fig.~\ref{fig:sota_rd_curves} shows R-D curves across datasets and LIC models, where our proposed two quantization strategies closely match full-precision baselines with minimal degradation over 0.1--1.0 bpp. Table~\ref{tab:rd_performance} reports fixed-bit-width quantization achieving near-optimal BD-Rate losses (Q-Cheng: 1.60\%, Q-Ballé: 5.01\%) compared to FMPQ benchmarks (1.30\%, 7.50\%). Dynamic bit-width quantization improves speedup (4.61$\times$, 5.17$\times$, 4.55$\times$ across models, surpassing 4$\times$) with slightly higher BD-Rate loss. These results confirm that DynaQuant's content-aware quantization and dynamic bit-width optimization enable efficient LIC deployment on resource-constrained devices.

\begin{table}[ht]
  \setlength{\tabcolsep}{4pt} 
  \centering
  \caption{General ablation study of our two quantization strategies: DPA (fixed bit-width, INT8/INT6) and DPA-DQ (dynamic bit-width quantization, DBWS), compared to Vanilla and PAMS quantization.} 
  \label{tab:total_ablation}
  \begin{tabular}{lccccc}
    \toprule
    Method & {Bitwidth} & {bpp} & {PSNR} & {R-D loss} \\
    \midrule
    Vanilla & 32 & 0.831 & 36.91 & 1.52 \\
    \midrule
    \multicolumn{5}{c}{\textit{Fixed Bit-width Methods}} \\
    \midrule
    +PAMS & 8 & 0.83 & 36.185 & {1.64} \\
    +{DPA[INT8]} (Ours) & 8 & 0.828 &  36.649 & \bfseries 1.56 \\
    +{DPA[INT6]} (Ours) & 6 &  0.827 & 35.664 & 1.74 \\
    \midrule
    \multicolumn{5}{c}{\textit{Dynamic bit-width Methods}} \\
    \midrule
    +PAMS-DQ & 6.85 & 0.892 & 30.262 & 4.28 \\
    +{DPA-DQ} (Ours) & 6.42 & 0.838 &  36.636 & \bfseries 1.57 \\
    +{DPA-DQ} (Ours)* &  6.02 & 0.882 & 36.231 & 1.68 \\
    \bottomrule
  \end{tabular}
\end{table}

\begin{table}[ht]
\setlength{\tabcolsep}{6pt}
\centering
\caption{Ablation study of our proposed fixed bit-width quantization strategy, DPA.}
\label{tab:DPA_ablation}
\begin{tabular}{ccccccc}
\toprule
$s$ & $z$ & $g(x)$ & Bitwidth & bpp & PSNR & R-D Loss \\
\midrule
\checkmark & \checkmark & \checkmark & 8 &  0.828 & \bfseries 36.649 & \bfseries 1.56 \\
× & \checkmark & \checkmark & 8 & 0.83 & 36.185 & 1.65 \\
\checkmark & × & \checkmark & 8 & 0.824 & 36.323 & 1.58 \\
\checkmark & \checkmark & × & 8 & 0.842 & 36.288 & 1.63 \\
\bottomrule
\end{tabular}
\end{table}

\begin{table}[htbp]
\setlength{\tabcolsep}{4pt}
\centering
\caption{Ablation study of our proposed dynamic bit-width quantization strategy (DBWS) with different bitwidth choices.}
\label{tab:bitwidth_ablation}
\begin{tabular}{lcccc}
\toprule
{Bitwidth Choice} & {Avg Bitwidth} & {bpp} & {PSNR} & {R-D Loss} \\
\midrule
\{4,6,8\} & 5.47 & 0.866 & 36.432 & 1.64 \\
\{6,8,10\} & 6.42 & 0.838 & 36.636 & 1.57 \\
\bottomrule
\end{tabular}
\end{table}

\begin{figure*}[ht]
  \centering
  \includegraphics[width=\textwidth]{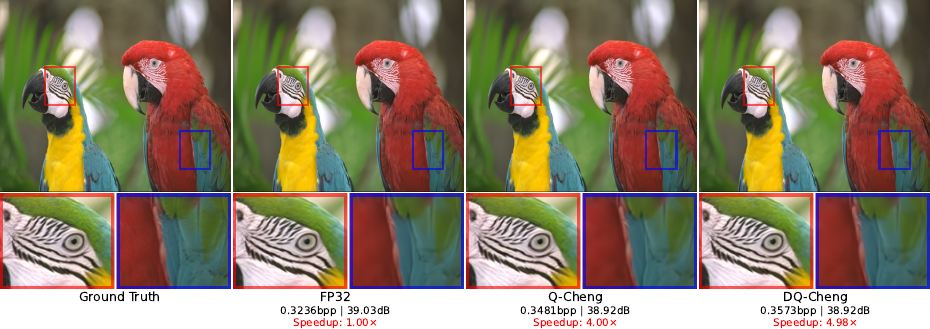} 
  \caption{Qualitative comparison on a Kodak image (e.g., ``\textit{kodim23}") using the base model {Cheng2020}. (a) Full-size images (first row). (b) Zoomed-in regions highlighting texture/edge details (second row). Methods: From left to right—Ground Truth, 32-bit full-precision,  Application of our Dynamic Parameter Adaptation (Q-Cheng), and application of a Dynamic Bit-Width Selector (DQ-Cheng). Metrics include bpp, PSNR, and speedup. Best viewed digitally and zoomed in.}
  \label{fig:qualitative_results}
\end{figure*}

\subsection{Ablation Studies}
\label{subsec:ablation_studies}
\paragraph{General Ablation}
Table~\ref{tab:total_ablation} presents the results of our general ablation study. \textbf{First}, our DPA method outperforms the classical 8-bit PAMS quantization, achieving an approximate bpp of 0.83 (ranging from 0.827 to 0.828), the highest PSNR (36.649), and the lowest R-D loss (1.56), confirming the effectiveness of our fixed-bit-width quantization approach. \textbf{Second}, integrating our dynamic bit-width method (DQ) with DPA reduces the average bitwidth from 8 to 6.42 while maintaining a competitive PSNR (36.636) and R-D loss (1.57), underscoring the efficiency of dynamic optimization. \textbf{Third}, the combination of DPA and DQ demonstrates a synergistic effect, outperforming the sum of individual contributions (i.e., DPA-DQ exceeds PAMS-DQ and simple DPA with INT6), with an optimized bitwidth of 6.02, a PSNR of 36.231, and an R-D loss of 1.68. This suggests that the integrated approach enhances performance beyond additive gains, validating the efficacy of each component and the superior impact of their combined use.
\paragraph{Ablation of Dynamic Parameter Adaptation }
Table~\ref{tab:DPA_ablation} shows that removing any DPA component degrades performance (e.g., excluding $s$ reduces PSNR from 36.649 to 36.185 and increases R-D loss from 1.56 to 1.65). These results highlight the importance of each component—learnable $s$, $z$, and $g(x)$—in improving quantization accuracy and preserving compression quality.

\paragraph{Ablation of Dynamic Bit-Width Selector}
Table~\ref{tab:bitwidth_ablation} shows that the dynamic bit-width strategy (DBWS) remains effective across different bit-width sets. The $\{6,8,10\}$ set offers a better trade-off between efficiency and fidelity, while $\{4,6,8\}$ achieves lower average bit-width at the cost of reduced PSNR.

\subsection{Qualitative Results}
\label{subsec:qualitative_results}
\paragraph{Visualization Results}
Fig.~\ref{fig:qualitative_results} demonstrates that our proposed quantization strategies (e.g., Q-Cheng and DQ-Cheng) achieve significant speedups of 4.00$\times$ and 4.98$\times$ respectively, despite slight bpp increases, while maintaining near 39~dB visual quality comparable to the 32-bit full-precision baseline.

\begin{figure}[t]
    \centering    \includegraphics[width=1\linewidth]{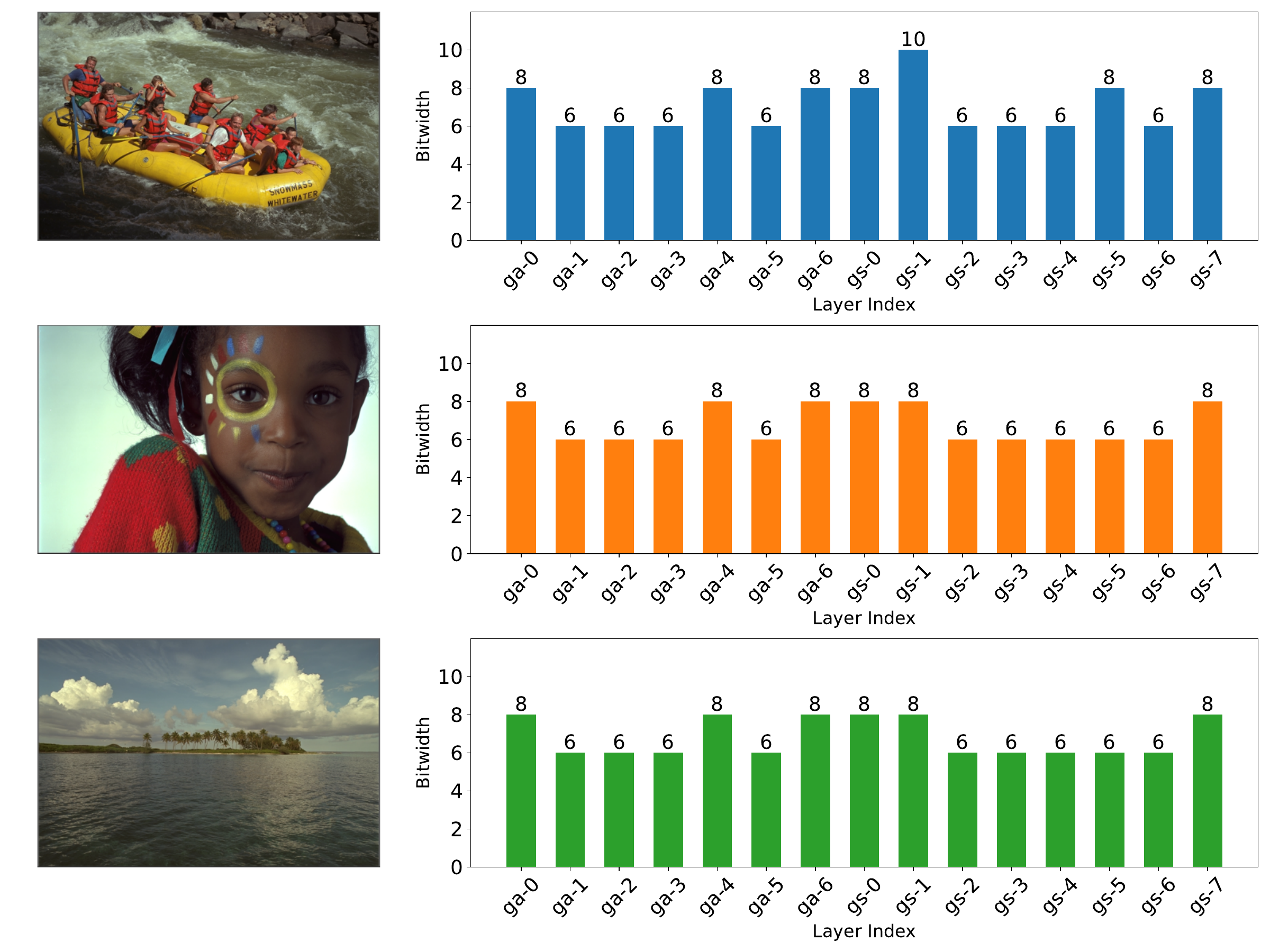}
    \caption{Learned bit-width allocation across network layers based on {Cheng2020}. Left: \textit{Kodim14/15/16} test images (top to bottom). Right: Optimized bit-widths automatically assigned to each layer.}
    \label{fig:cheng_layer_distribution}
\end{figure}

\paragraph{Bit-width Selection Results}
Fig.~\ref{fig:cheng_layer_distribution} showcases dynamic bit-width allocation across images and layers. For example, \textit{Kodim14} with more detailed texture assigns a 10-bit width to its gs-1 layer, exceeding the 8-bit allocation in other images. Edge layers (e.g., ga-0, ga-6, gs-1) exhibit higher precision than intermediate layers, indicating bottleneck layers require increased bit-width. These findings confirm that our Dynamic Bit-width Quantization adapts precision to data complexity and network structure.

\section{Conclusion}
\label{sec:conclusion}

In this work, we presented DynaQuant, a dynamic mixed-precision quantization framework for LIC that jointly adapts quantization parameters and layer-wise bit-widths in a content-aware manner. By employing a distance-aware gradient modulator to train learnable quantization parameters, the method provides more informative gradients and mitigates performance degradation. A differentiable dynamic bit-width selector further learns to allocate optimal precision to each network layer by jointly optimizing the rate–distortion loss with a differentiable mixed-precision allocation cost. Extensive experiments show that DynaQuant achieves R–D performance close to that of full-precision models while significantly reducing computational complexity and model size, yielding speedups of up to 5.17$\times$. \par
Although a performance trade-off remains at maximum acceleration, future work may refine this balance by incorporating hardware-specific latency models into the optimization objective.

\section*{Acknowledgments}
This work was supported in part by the National Natural Science Foundation of China (Grants 62571160 and 62472124), the Shenzhen Colleges and Universities Stable Support Program (Grant GXWD20220811170130002), the Engineering Technology R$\&$D Center of Guangdong Provincial Universities (Grant 2024GCZX004), and the Major Project of Guangdong Basic and Applied Basic Research (Grant 2023B0303000010).

{
\small
\bibliographystyle{aaai2026.bst}
\bibliography{aaai2026}
}

\counterwithin{figure}{section}  
\counterwithin{table}{section}   
\counterwithin{equation}{section} 
\clearpage
\setcounter{page}{1}

\onecolumn
\section*{Appendix}
\label{sec:appendix}
This appendix provides comprehensive technical details on the following aspects:
\begin{enumerate}
    \item The architectural implementation of the Bit-Width Selector module, including detailed hyperparameter configurations of convolutional and pooling layers, network topology, and data flow propagation paths;
    \item Visualization results of bit-width allocation strategies of the proposed Bit-Width Selector module across different neural network architectures, demonstrating the quantization bit-width selection distribution for each functional module.
\end{enumerate}

\renewcommand{\thesubsection}{A\arabic{subsection}} 
\renewcommand{\thefigure}{A\arabic{figure}}    
\renewcommand{\thetable}{A\arabic{table}}      
\renewcommand{\theequation}{A\arabic{equation}} 
\setcounter{figure}{0}
\setcounter{table}{0}
\setcounter{equation}{0}

\subsection{A.1 DBWS  Architecture Details}
\subsubsection{A.1.1 Encoder-Decoder DBWS Configuration}
For the encoder and decoder, each is equipped with its own independent dynamic bit-width selector, but both adopt identical network architectures. This symmetric architectural design ensures that the encoder and decoder can generate consistent bit-width selection strategies, eliminating the need to transmit additional bit-width configuration information during image transmission and significantly reducing communication overhead. In the following descriptions, the number of quantized layers in the encoder and decoder is uniformly denoted as $BL$.

\subsubsection{A.1.2 DBWS Design}
\begin{figure}[H] 
    \centering
    \includegraphics[width=1\linewidth]{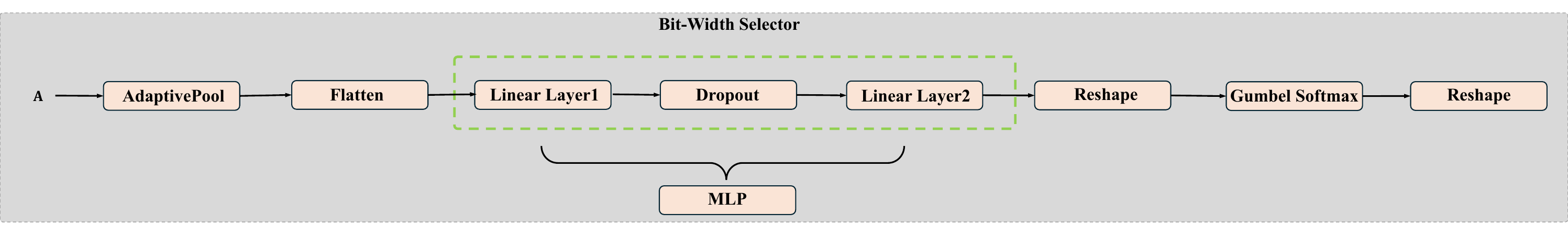}
    \caption{Detailed Architecture of DBWS}
    \label{fig:bws_architecture}
\end{figure}

As illustrated in Figure~\ref{fig:bws_architecture}, the DBWS module adopts a hierarchical architecture with parameter configurations detailed in Table~\ref{tab:bws_implementation}. The module comprises six core components:

\begin{enumerate}[label=(\arabic*), itemsep=0pt, parsep=0pt]
    \item \textbf{Adaptive Pooling Layer:} pools input features $A$ to a fixed size $(5 \times 5)$, ensuring generalizability for different input resolutions;
    \item \textbf{Flatten Layer:} converting feature maps to one-dimensional vectors;
    \item \textbf{Multi-Layer Perceptron (MLP):} consisting of two linear layers and one dropout layer, implementing feature transformation and dimensionality reduction from $N \times 25$ to $128$, then outputting $\text{num\_bits} \times BL $ dimensional feature vectors;
    \item \textbf{Reshape Layer:} reconstructing outputs to $(B \times  BL, \text{num\_bits})$ shape;
    \item \textbf{Gumbel Softmax Layer:} performing differentiable discrete sampling with hard quantization strategy ($\tau = 1$, \texttt{hard=True}) to output definitive bit-width configurations;
    \item \textbf{Final Reshape:} reshaping to $(B,  BL, \text{num\_bits})$ to provide one-hot encoded bit-width selections for $BL$ quantization blocks.
\end{enumerate}

The DBWS module achieves adaptive optimal bit-width selection based on input features through end-to-end learning, balancing compression efficiency and reconstruction quality.\\
\textbf{Notice:} In our experimental setup and training process, the dynamic bit-width selector is applied exclusively to the main encoder-decoder module, while the hyperencoder adopts a uniform 8-bit fixed quantization strategy. This design decision is based on the following experimental analysis: the hyperencoder module is characterized by simple structure and fewer parameters, making further quantization compression yield limited benefits, with negative returns observed in some cases; meanwhile, each layer of this module exhibits high sensitivity to quantization errors, where low bit-width quantization easily leads to significant performance degradation; furthermore, given the relatively small computational load of the hyperencoder itself, introducing complex dynamic quantization strategies would add unnecessary system overhead with limited performance improvement. Therefore, adopting a uniform 8-bit quantization strategy for the hyperencoder represents the optimal choice, ensuring both module performance stability and avoiding excessive system architecture complexity.

\begin{table}[H] 
\centering

\caption{Detailed Parameter Configuration of DBWS}
\label{tab:bws_implementation}
\renewcommand{\arraystretch}{1.7}
\begin{tabular}{@{}llllll@{}}
\toprule
\textbf{Layer} & \textbf{Input Shape} & \textbf{Output Shape} & \textbf{Parameters} & \textbf{Key Settings} \\
\midrule
AdaptivePool & (B, N, H, W) & (B, N, 5, 5) & 0 & output\_size=(5,5) \\
Linear & (B, N×25) & (B, 128) & (N×25+1)×128 & in\_feature=N×5×5,out\_feature=128\\
Dropout & (B, 128) & (B, 128) & 0 & p=0.2 \\
Linear & (B, 128) & (B×BL,num\_bits) & (128+1)×(num\_bits×BL) & in\_feature=128,out\_feature= BL×num\_bits\\
GumbelSoftmax & (B×BL,num\_bits) & (B×BL,num\_bits) & 0 & tau=1,hard=True\\
\bottomrule
\end{tabular}
\end{table}

\subsection{A.2 DBWS Input Data Selection Strategy}
\begin{figure}[H] 
    \centering
    \includegraphics[width=1\linewidth]{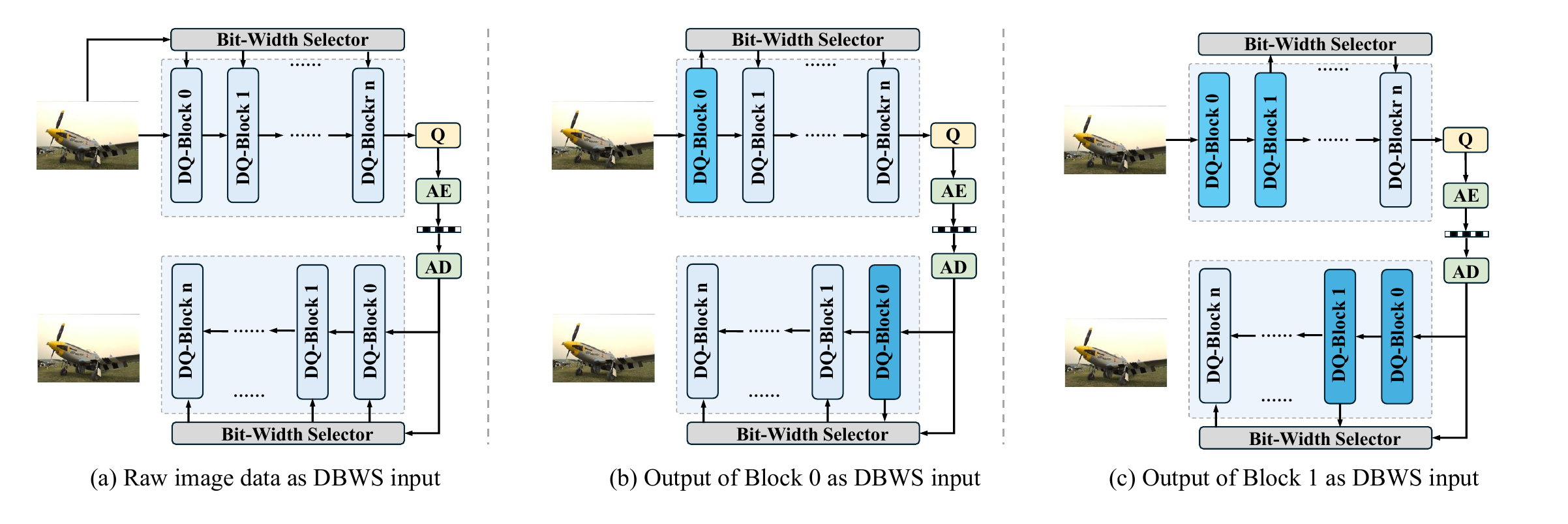}
    \caption{
        Dark blue DQ-Block modules denote fixed 8-bit quantization, while light blue modules represent adaptive quantization blocks.}
    \label{fig:compare}
\end{figure}

In our experimental framework, the Dynamic Bit-Width Selector (DBWS) for the main encoder-decoder adopts a symmetric configuration design, where the encoder and decoder respectively utilize the output features from their corresponding first modules as inputs to their respective DBWS. The core design of this configuration strategy is as follows:
First, the first processing modules of both encoder and decoder employ fixed $8$-bit quantization to provide a stable feature foundation for subsequent bit-width selection; second, the encoder DBWS processes output features from the encoder's first module, while the decoder DBWS processes output features from the decoder's first module; finally, based on the respective DBWS outputs, all subsequent modules (2nd to $BL$th) in the encoder-decoder employ adaptive bit-width selection strategies.
This symmetric configuration design maximizes quantization flexibility while ensuring computational efficiency, enabling most processing modules in the encoder-decoder to benefit from dynamic quantization advantages.
\\
As illustrated in Figure~\ref{fig:compare}, we present a comparative analysis of different layer outputs serving as DBWS inputs. After comprehensive trade-off analysis between dynamic quantization flexibility and computational complexity, we ultimately determine to adopt the output features from the respective first modules of the encoder-decoder as the corresponding DBWS input configuration.

\subsection{A.3 Cross-Architecture Results}
\begin{figure}[H]
    \centering    \includegraphics[width=1\linewidth]{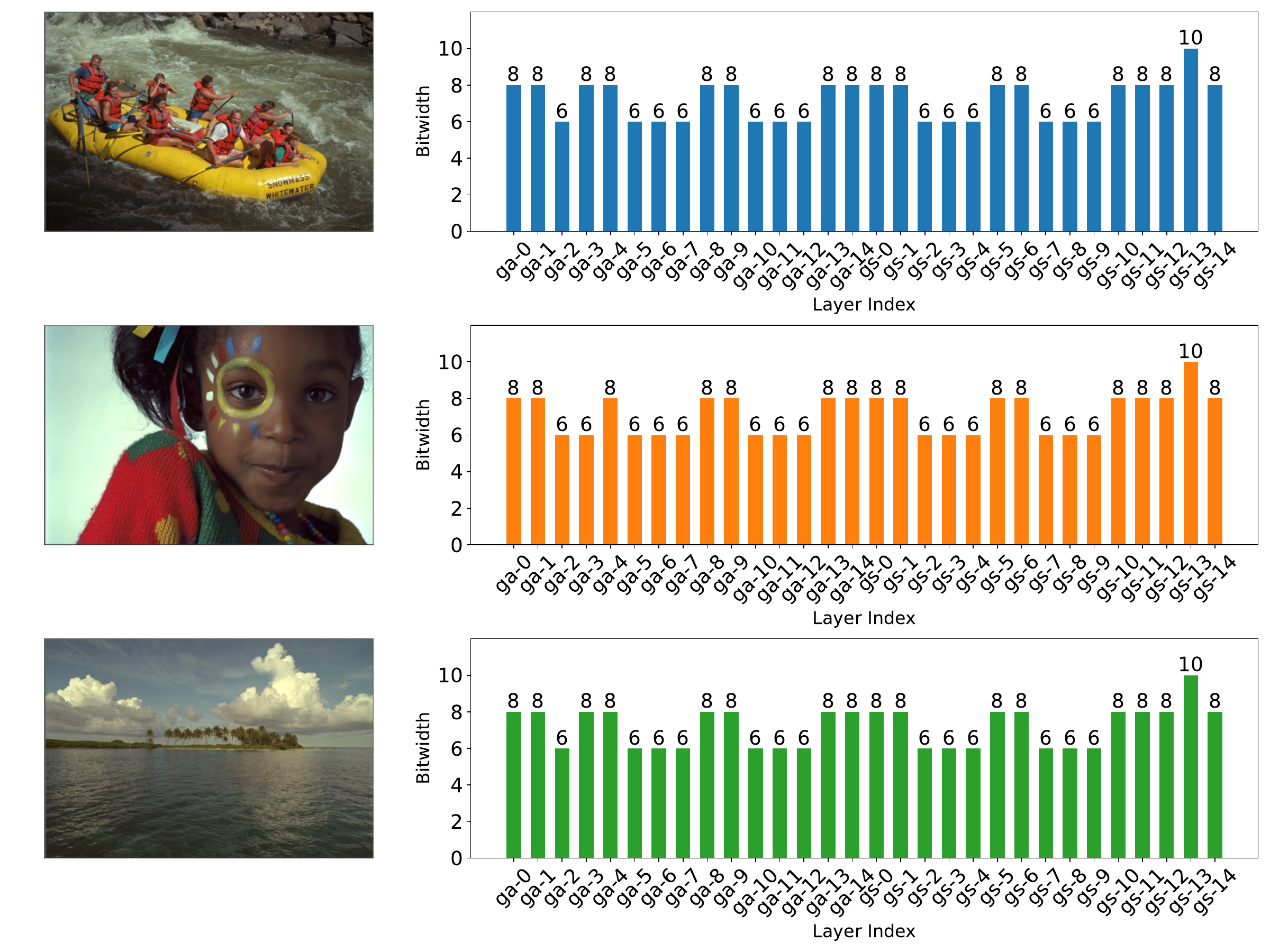}
    \caption{Learned bit-width allocation across network layers based on {ELIC}. Left: Kodim14/15/16 test images (top to bottom). Right: Optimized bit-widths automatically assigned to each layer.}
    \label{fig:ELIC_layer_distribution}
\end{figure}

\begin{figure}[H]
    \centering    \includegraphics[width=1\linewidth]{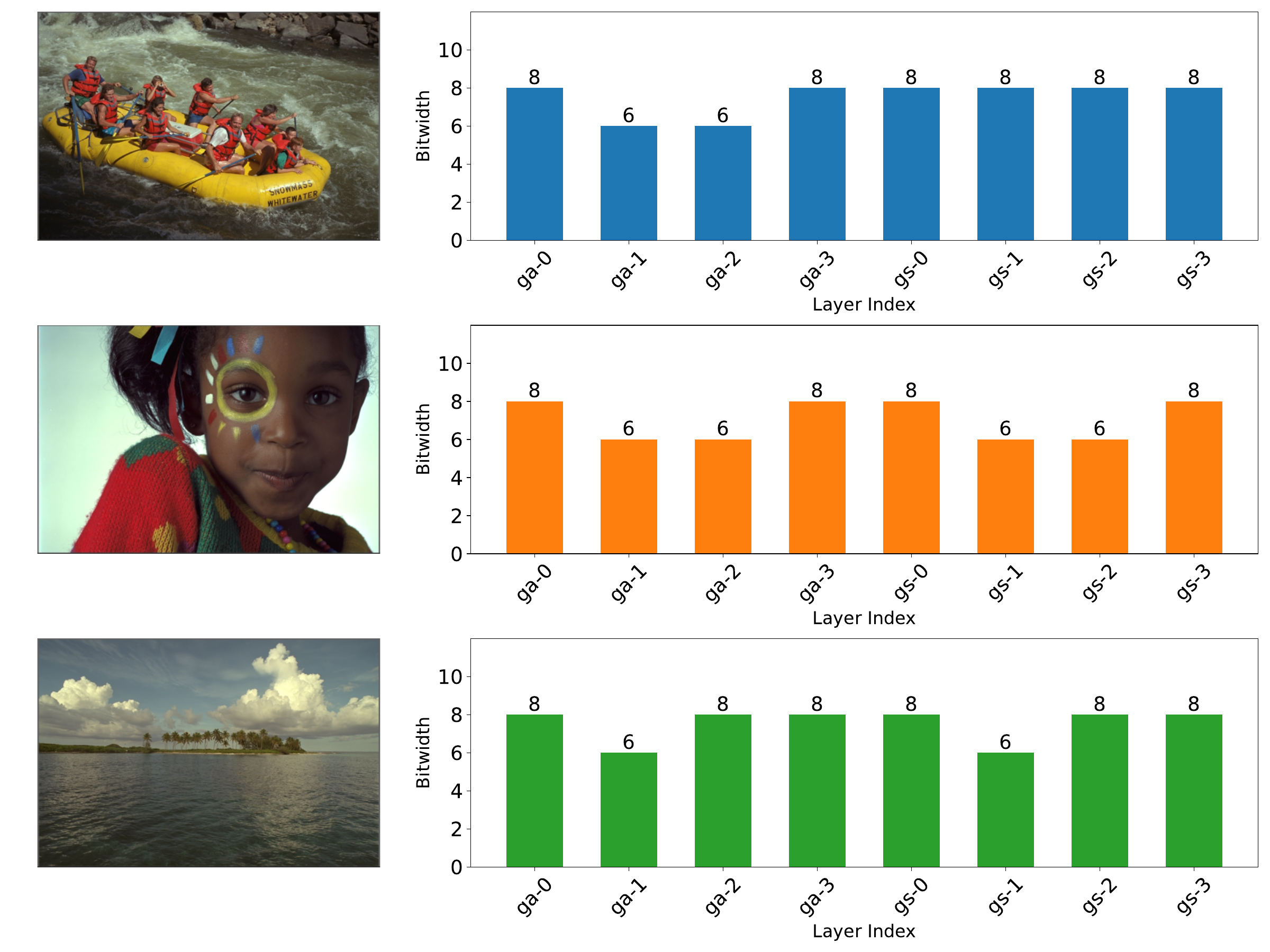}
    \caption{Learned bit-width allocation across network layers based on { Ballé}. Left: Kodim14/15/16 test images (top to bottom). Right: Optimized bit-widths automatically assigned to each layer.}
    \label{fig:Balle_layer_distribution}
\end{figure}

To validate the generalizability of our Dynamic Bit-Width Selector (DBWS) Selector (DBWS), we extend the evaluation to two additional neural compression architectures: Ballé and ELIC. Figure~\ref{fig:ELIC_layer_distribution} and Figure~\ref{fig:Balle_layer_distribution}  presents the bit-width allocation patterns across different layers and test images for these alternative architectures.

\end{document}